\documentclass[11pt,aps,preprint,epsfig,showpacs]{revtex4}
\usepackage{graphicx}
\usepackage{epsfig}
\begin{document}
\title{Effects of Eye-phase in DNA unzipping} 
\author{Debaprasad Giri$^\dagger$ and Sanjay Kumar} 
\affiliation{Department of Physics, Banaras Hindu University,
Varanasi 221 005, India \\
Max-Planck Institute for the Physics of Complex Systems, Noethnitzer, 
01187 Dresden, Germany\\
$^\dagger$Physics Section, MMV, Banaras Hindu University,
Varanasi 221 005, India} 

\begin{abstract}
The onset of an ``eye-phase" and its role during the DNA unzipping
is studied when a force is applied to the interior of the chain.
The directionality of the hydrogen bond introduced here shows
oscillations in force-extension curve similar to a "saw-tooth" 
kind of oscillations seen in the protein unfolding experiments.  
The effects of intermediates (hairpins) and stacking energies 
on the melting profile have also been discussed. 
\end{abstract}
\maketitle

Molecular interactions play a key role in living organisms.  
Recent advances in experimental techniques have allowed 
nanomanipulation in single biological molecule and made 
possible to measure these interactions \cite{sinmol}. 
The aim is to exert a force in the pN range by optical 
tweezers, atomic force microscopy, etc and characterize
the molecular, elastic, structural and functional properties 
of bio-molecules \cite{cof,ubm}. In typical experiments of 
double stranded DNA (dsDNA) unzipping, a force is applied 
to the ends of the chain (Fig. 1a) and one studies the 
force-extension curve which shows the elastic properties 
and the gross features that reflect the local `G-C' vs `A-T' 
content along the sequence \cite{smh,mgz,chl,rief,bloom}. Moreover, the 
force-temperature diagram shows, below the melting temperature, 
that the over-stretching force \cite{bloom} decreases nearly 
linearly with temperature.

Theoretically, DNA unzipping may be studied in different ensembles 
\cite{busta} depending on the experiments. For example, Atomic force 
microscopes ({\bf AFM}) work in constant distance ensemble ({\bf CDE}) 
while magnetic bead uses the concept of constant force ensemble ({\bf CFE}).
The prediction of unzipping transition based on interacting 
Gaussian chains \cite{bhat99}  
 raised a lot interest 
and now results are available from dynamical approach \cite{seba},
exact solutions of lattice models \cite{trieste,kbs}, simple models of 
quenched-averaged DNA \cite{nelson,lam}, numerical simulations and 
scaling analysis \cite{nsys,chen}. Recently for a model of interacting 
polymers where any monomer of one chain can interact with any monomer 
of the other chain (we call it {\bf model A}), the role of an 
intermediate entropy-stabilized phase was recognized and a force-induced 
triple point \cite{kgbhat} in a force-temperature plane was established.  

In most of the models studied for dsDNA
\cite{trieste,kbs,nelson,lam,nsys,baiesi}, a monomer $i$ of one strand can 
only interact with the $i$-th monomer of the other strand,  which is 
similar to the models of DNA (we call it {\bf model B})
proposed earlier 
by Poland and Scheraga \cite{polsch}. 
These models do not take into 
account the directional nature of the hydrogen bond  
and underestimate the entropy by restricting the formation 
of hydrogen bonds. Thus these models may only give a limited picture of 
the unzipping transition and do not allow to study the effect of 
intermediate states 
\cite{danilow}.

\begin{figure}[h]
\includegraphics[height=2.2in,width=2.4in]{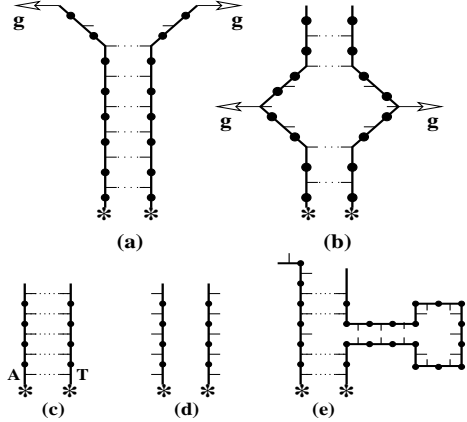}
\caption{The schematic representation of DNA unzipping by force ($g$) 
applied at (a) one end of the two strands; (b) at the interior of the 
strands. Figs. (c-e) represent the possible conformations of model C. 
Here (c) represents the completely zipped state, (d) a non pairing 
configuration and (e) a partial bound state with hairpin loop.}
\label{fig-1}
\end{figure}

So far, most of the theories of DNA unzipping have focused on
the thermodynamic limit, and therefore, consist of a few parameters
typically in the form of an effective base-pair interaction in a
simplified way. In all the single molecule experiments, a finite 
size chain is used and hence no ``true phase transition" can  
in principle be observed. Still, the ``phase transition" observed
in such experiments may be considered as real if the length of the
chain exceeds the characteristic correlation lengths. It is now
becoming possible to go to the other limit of studying shorter
segments ($\sim 10$ base pair) at coarse grained level \cite{sosg,lfenb}.
The purpose of this letter is to provide exact results of a semi-microscopic 
model of short chains by incorporating the directional nature of 
hydrogen bonds and then propose a method to study the effect of molecular 
interactions right at the individual base pair level and their role on melting 
profile. 

In the following, we adopt a more realistic model of DNA, 
which may be defined in any dimension \cite{kgs} (herein after
we call it {\bf model C}). A similar model has been used in 
Ref. 24 in the context of relative stabilities of 
DNA hairpin structures.  The model takes care of important 
shortcomings of model B and also incorporates some 
additional features like existence of intermediate states, effects
of stacking energy, excluded volume properties of nucleotides and 
the directional nature of hydrogen bonds.  

Although the importance of bubble formation during thermal melting 
has been recognized, no attempt has been made so far to experimentally
explore the phase diagram when a force is applied to the interior 
of the chain as shown in Fig 1b. Such situations occur in many 
biological processes, for example, during gene-expression, 
RNA forms bubbles or ``eye-type" conformations on DNA.  
Therefore, we consider two cases: ($i$) 
force has been applied at the end of the chain (Fig. 1a,`END' case
or $Y$ case), and ($ii$) 
at the middle  of the chain (Fig. 1b, `MID' case).  
The contribution to energy by this force, $g$, is $-2g x$, where $2x$ 
is the absolute distance in the $x$-direction between the two strands 
at the point of application of the force.

We model the two strands ({\it e.g.} A-T) of a homopolymer DNA by two mutually-
attracting-self-avoiding walks ({\bf MASAWs}) on a square
lattice as shown in Figs. 1(c-e).
The bases are associated with the link 
between two monomers of a chain as depicted in Fig. 1. In one strand the 
bases point towards the right while on the other they are on the left, 
as one traverses the chains sequentially. We associate a contact energy 
$-\epsilon$ (effective base pair interaction) with each pairing between 
complementary strands only if the bases are nearest neighbors (short range 
nature of the hydrogen bond) and approach each other directly without 
the strands coming in between [Fig. 1c].  
Fig. 1e shows the possibility
of formation of hairpin (which is not possible in model B) in a single strand of DNA. 
However, in this case, non-native contribution has been taken into 
account but no apparent weight has been assigned to stem as it is 
made up of same nucleotides.

The partitions function ($Z_N$) of the system under consideration can be 
written as a sum over all possible configurations of {\bf MASAW}s {\it i.e.}
$\sum_{m,x} C(m,x) \exp(\beta m \epsilon) \exp(\beta g x)$, where $\beta =1/k_BT$ 
is the inverse temperature, $k_B$ being the Boltzmann constant. $C(m,x)$ is 
the number of distinct conformations walks of length $2N$ having $m$ number
of intact base pairs whose end (or mid) points are at a distance $x$ apart.
We have obtained $C(m,x)$ for $N \le 16$ and  analyzed the partition function
using exact enumeration and series analysis technique \cite{ratio,exenu}.
We prefer this technique because it can predict various phases of the system  
\cite{maren} quite effectively and the scaling corrections can be correctly 
taken into account by a suitable extrapolation methods \cite{ratio,exenu}.
To achieve the same accuracy in Monte Carlo, a chain of two orders of
magnitude larger than in the exact enumeration method is required 
\cite{ykg}. We set $\epsilon/k_B = 1$ and calculate the reduced 
free energy per base pair  from the relation
$G(T,g)= \lim_{N \rightarrow \infty} \frac{1}{N} \log Z(T,g) = \log \mu(T,g)$
\cite{exenu}.
The limit $N \rightarrow \infty$ is achieved by using the ratio method 
\cite{ratio} for extrapolation. The transition point  
can be obtained from the plot of $G(T,g)$ versus 
$T$ or from the peak value of $\frac{\partial^2 G}{\partial(1/T)^2}$. 

A force-temperature ($g-T$) phase diagram of model B and C for end and mid 
case is shown in Fig. 2. The qualitative features of the phase diagram 
obtained here may be compared with experiments \cite{bloom}. The phase 
boundary separates the zipped and the unzipped state.  At $T=0$, the 
critical force can  be found from a simple analytic calculation and is 
equal to $0.5$ \cite{kgbhat} that is in agreement with the one from 
Fig. 2. 

\begin{figure}
\includegraphics[height=1.8in,width=1.8in]{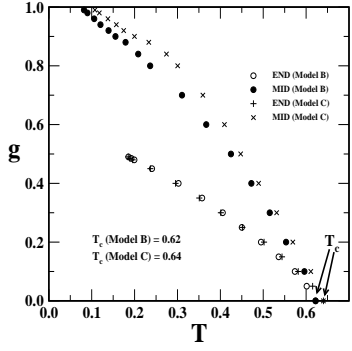}
\caption{Variation of critical force ($g_c(T)$) (END and MID case) as a 
function of temperature ($T$) for model B and C.}
\label{fig-2}
\end{figure}

Experimentally, the melting profiles are obtained by monitoring the change
in the UV absorbence with temperature which provides the information about
the fraction of open base pairs and the melting temperature is defined when
half of the total base pairs get open \cite{melt}. 
Another quantity of experimental interest is to monitor and measure 
the end separation by varying the force. We calculate these quantities from 
the expressions $<m>=\sum m C(m,x) \exp(\beta m \epsilon) \exp(\beta g x) / Z_N$
and 
$<x>=\sum x C(m,x) \exp(\beta m \epsilon) \exp(\beta g x) / Z_N$  
and plot their variations with temperature and force in Figs. 3 and 4, 
respectively.  It is evident from Fig. 3 that, with increasing temperature 
(or force), the number of intact base pairs decreases and there is a sharp 
transition from a zipped state to the unzipped state. 

\begin{figure}
\centerline{ \epsfig{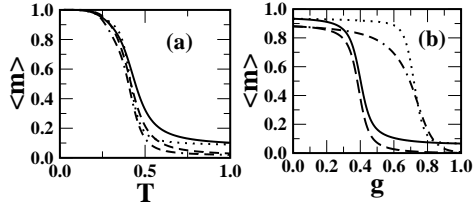}}
\caption{Variation of scaled $<m>$ (a) with temperature ($T$) at constant force 
$g=0.25$ (END) and $0.50$ (MID); (b) with force ($g$) at constant $T=0.3$. 
Here solid and dashed lines represent the END case, while dot and dot-dashed  
line represent the MID case for model B and C respectively.}
\label{fig-3}
\end{figure}

\begin{figure}
\vspace {.1in}
\centerline{\epsfig{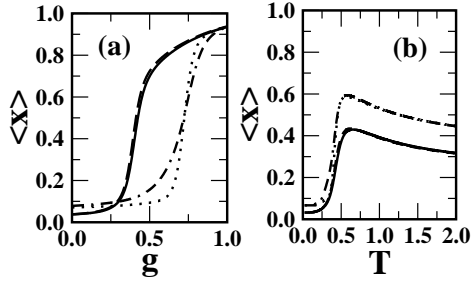}}
\caption{Plot of scaled $<x>$ (a) with force ($g$) at constant $T=0.3$ and (b) with
temperature ($T$) at constant $g=0.25$ (END) and $0.5$ (MID). The lines have 
same meaning as of Fig. 3.}
\label{fig-4}
\end{figure}
\vspace{.1in}
Remarkably, to break the same amount of base pairs at low temperature 
one requires almost double the force if it is applied in the middle 
of the chain, rather than the end (Fig. 3b) consistent with the exact 
results \cite{kbs}.  
Near the melting temperature, 
fluctuation dominates and less than double the force is required for 
unzipping the chain from the middle. 
\begin{figure}
\vspace {.2in}
\centerline{\epsfig{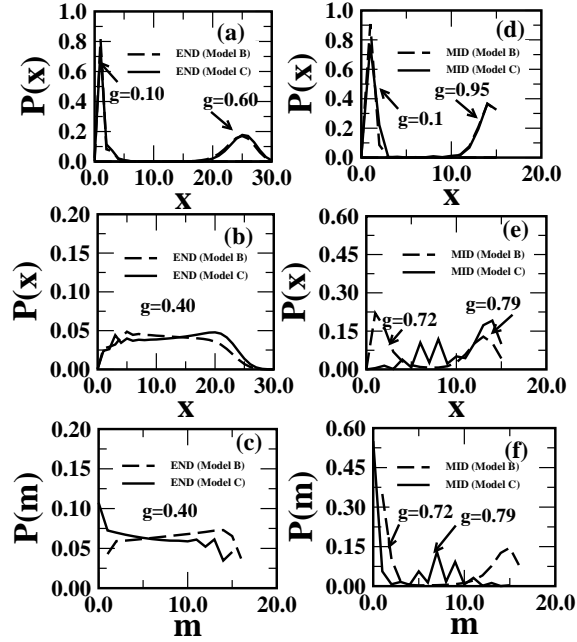}}
\caption{Figs. (a-f) represent the $P(x)$ and $P(m)$ of END case (a-c) 
and MID case (d-f) for different forces at constant T=0.3. Fig. 5e shows
the signature of ``eye-phase" of even widths in the form of oscillation 
for the mid case in model C but absent in model B.   
}
\label{fig-5}
\vspace {-0.2in}
\end{figure}

Another interesting observation is the variation of average elongation 
with force which shows a monotonically increasing trend (Fig. 4a) 
at constant temperature and approaches unity. Meanwhile the variation 
of extension at constant force shows a sharp rise with temperature 
(Fig. 4b) and then a slight decline to approach a value below unity. 
At constant temperature, there is a transition from the zipped to the 
unzipped (``rod-like") state.  Keeping force constant, when temperature 
is varied, there is a transition and the chain acquires conformations 
close to the rod-like states. As this temperature is still low, with 
further increase in temperature, the entropy of the system increases and 
the chain acquires coil-like state at higher temperature and thus average 
distance decreases.

We also study the probability distribution curves $P(x)$ with $x$ and 
$P(m)$ with $m$ for model B and C defined by
$P(x)= \sum_{m}
 C_N (m,x) \exp(\beta m \epsilon)  \exp(\beta g x)/Z_N$ and 
$P(m)= \sum_{x}
 C_N (m,x) \exp(\beta m \epsilon) \exp(\beta g x)/ Z_N$ respectively.
In Figs. 5(a-f) we have shown $P(x)$ and $P(m)$ for different 
values of force and a fixed temperature $T=0.3$. The $x$-component 
of the distribution function gives information about the states of dsDNA. 
The maxima of $P(x)$ at $x \approx 0$ correspond to the zipped 
state for a given set with $g=0.1$ and $T=0.3$. For small forces, 
thermal fluctuations are too weak to unzip the strand and the DNA 
remains in  the zipped state. This is being reflected in the structure-less 
distribution function with a well defined peak at the most likely 
value of the extension (Figs. 5a and 5d).  For both the models at 
higher force and at the same temperature ($g=0.6$ for end and $0.95$
for mid respectively at $T=0.3$), thermal fluctuations have been 
suppressed by a strong force and strands are found 
in the segregated ``rod-like" state with more or less an identical 
distribution peaked at maximum extension.  
However, striking differences are observed in the probability
distribution curves for model B and C at intermediate forces below
transition line ($g-T$ plane) when the force is applied in
the middle. No such differences are observed for the end case. 
For the mid case, the probability distribution curve of model C 
shows strong oscillations, whereas for model B no such oscillations 
are observed. 
This indicates that the model B may be described by two state models. 
However, model C shows that certain intermediates states (``eye-shape" 
of even width) are more favorable than others. 

\begin{figure}
\centerline{\epsfig{file=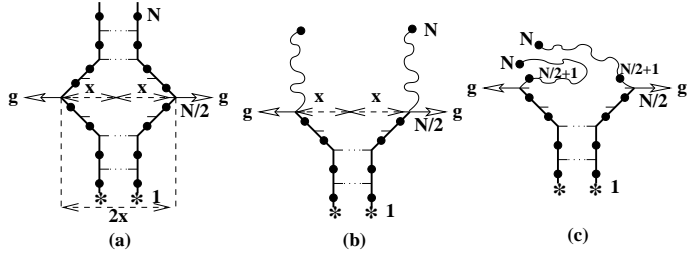, scale=0.4}}
\caption{Schematic diagrams showing the different conformations 
keeping $2x$ $(\le N^\nu)$ distance constant. For model B, (b) and (c) has
same weight while for model C, (c) has more weight than (b).}
\label{fig-6}
\vspace {-.2in}
\end{figure}

It is interesting to note that the $g-T$ phase diagram of model B and C
are almost identical without any re-entrance at low temperature. This
may be because of the fact that the energy of the ground state and the 
unbound state for both models are the same. At the center point of the 
`Y' (end case), the phase boundary is determined by a balance of the 
net force -$2gx$ and the unzipping potential $-\epsilon m$ with 
associated entropy.  In the thermodynamic limit, directionality of the 
hydrogen bonds, which appear in the form of entropy of the partial 
bound states does not play a crucial role in this balance. The absence of 
re-entrance is understood with the zero entropy of the ground state 
for both models. As shown recently by  Kapri {\it et al.} \cite{kbs} for mid case, 
in the {\bf CDE} there is a possibility of a coexistence 
region that is better thought of as an ``eye-phase" $\equiv$ two `Y' joined 
together. In this case, the separation at the point of 
application of force is smaller than the fully open case and thus 
such conformations statistically have more weight than the other conformations.

Since the bottom end is kept fixed, the top side of the strand 
may open due to thermal fluctuations (Fig. 6b) and form a partial bound 
state as shown in Fig. 6c which is more stable than Fig. 6b. 
Therefore, in 
model C, the half of the chain undergoes an unzipping transition while 
other half due to the non-native contacts, shows the combined effects 
of unzipping and slippage (shearing) transition. Thus in the 
model C, the transition  appears more smoother than the model B 
which can be seen in Fig. 3b.
The consecutive peaks in $P(x)$ vs $x$
curve (Fig. 5e) represent the ``eye-phase" of even widths which 
contribute most to the partition function. Because of 
native contacts, the contribution of the ``eye-shape" conformation is
significantly less and small thermal fluctuations are sufficient to
unzip the chain in model B.

We substantiate our arguments by extending calculation in {\bf CDE} also.  
The partition function in {\bf CDE} may be defined as $Z_N(x,T) = 
\sum_m \exp(\beta m \epsilon)$. The two ensemble are related by 
$Z_N(T,g) = \sum_x Z_N(x,T) \exp(\beta g x)$ \cite{kbs,nelson}. 
The free energy is  given by the relation $F_N(x,T)= -T \ln Z_N(x,T)$. 
In {\bf CFE} the average separation $<x>$ fluctuates while in {\bf CDE} 
one measures the average force to keep the separation constant given 
by the expression $<g> =\frac{\partial F_N (x,T)}{\partial x}$ at constant 
temperature \cite{busta}. The force-extension curve thus obtained is 
shown in Fig. 7 for model C which also shows oscillations for the `MID' 
case but constant for the `END' case. Though a somewhat similar effect has been
seen experimentally in molecules like Titin \cite{protein}, DNA has
not been probed so far.
The average of force obtained here ($0.2 \pm 0.05$ for `END' case 
and $0.35 \pm 0.05$ for `MID' case at $T=0.5$) also matches with the value 
shown in Fig. 2.  

\begin{figure}
\vspace {.2in}
\centerline{\epsfig{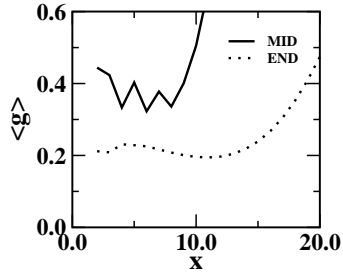}}
\caption{Plot of $<g>$ with distance ($x$) at constant $T=0.5$.}
\label{fig-7}
\vspace {-0.2in}
\end{figure}

The stacking energy in case of homopolymers gets adsorbed in the 
effective base pair interaction. To see this, we associate an 
additional energy between two consecutive  parallel base pairs (only 
possible in model C) and found that there is no qualitative change
in the phase diagram except shift in the transition temperature. 
If stacking energy is negative (attractive interaction) chain becomes
stiffer. However, it does not change the nature of oscillations as
observed in probability distribution curves for the `MID' case. 

The exact results on short chains of a new semi-microscopic dsDNA
that incorporates the directional nature of hydrogen
bond show unequivocal signature of an ``eye phase", without going to
the long chain limit. This happens for the case with a force acting
in the middle of the dsDNA. 
The variation of
elongation due to the force in different ensembles 
has different behaviour both qualitatively and
quantitatively. We anticipate that refinements in
high precision single molecule experiments will be able to verify
these predictions.

We thank  Y. Singh, S. M. Bhattacharjee and Jeff Chen for many 
fruitful discussions on the subject and University Grants Commission, India for the financial assistance.  One of us (SK) would like to 
acknowledge financial support from MPIPKS, Dresden, Germany.

\end{document}